\documentstyle[twocolumn,aps,prl]{revtex}
\begin{document}
\draft
\title{
Smooth-filamental transition of active tracer fields stirred by chaotic
advection}
\author{Zolt\'an Neufeld$^{1}$, Crist\'obal L\'opez$^{2}$ and
Peter H. Haynes$^{1}$}

\address{$^{1}$ Department of Applied Mathematics and Theoretical Physics,
University of Cambridge, Silver Street, Cambridge CB3 9EW, UK\\
$^{2}$ Instituto Mediterr\'aneo de Estudios Avanzados, IMEDEA (CSIC-UIB), 
07071 Palma de Mallorca, Spain}
\date{\today}
\maketitle

\begin{abstract}
The spatial distribution of interacting chemical fields is investigated
in the non-diffusive limit. The evolution of fluid parcels
is described by independent dynamical systems driven by chaotic
advection. The distribution can be filamental or smooth depending on
the relative strength of the dispersion due to chaotic advection
and the stability of the chemical dynamics. We give the condition
for the smooth-filamental transition
and relate the H\"older exponent of the filamental structure to
the Lyapunov exponents.
Theoretical findings are illustrated by numerical experiments.
\end{abstract}

The Lagrangian description of fluid flows, considered in the
framework of chaotic dynamical systems, has given important insight
into mixing and transport \cite{Ottino}.
In many situations of practical interest scalar fields 
describing some property of the fluid 
are not just simply advected by the flow,
but are {\em active} in the sense that they have their
own dynamics,
in general
coupled to the transport and mixing process. 
(In the following we will refer to this as the 'chemical'
dynamics of the system.) Typical examples of active fields
in mixing flows are concentrations
of reacting chemicals \cite{Epstein,Sokolov}
(in industrial processes or the atmosphere) 
and interacting biological
populations of microorganisms in a fluid
(e.g. plankton populations stirred by 
oceanic currents \cite{Abraham}). 
The spatial structure of such fields
often has complex filamental character \cite{Eduard}.
Previous work has investigated the temporal
evolution of reactions like $A+B \to C$ or $A+B \to 2B$ in
closed flows \cite{Sokolov,Metcalfe,Barcelona}
as an initial value problem. 
The absence of chemical sources in these cases
necessarily implies a 
homogeneous final state of the system.
The same reactions were also studied in 
open chaotic flows \cite{Tel}
where a stationary fractal distribution arises due to
the properties of the underlying (chaotic scattering-like)
advection dynamics \cite{Tel1}. 
Here we show 
that in the case of stable chemical dynamics (in a sense defined below)
and in the presence of chemical sources
persistent filamental fractal patterns can also arise in closed flows.

The governing equations for a set of $N$ interacting 'chemical' fields $C_i$ 
mixed by a flow ${\bf v}({\bf r},t)$, independent of
the chemical dynamics, can be written as
\begin{equation}
\frac{\partial {C_i}}{\partial t}+{\bf v}({\bf r},t) \cdot \nabla {C_i}=
{\cal F}_i[{C_1({\bf r}),..,C_N({\bf r})},{\bf r}],
\label{general}
\end{equation}
where $i=1,..,N$. The operators ${\cal F}_i$ 
in general contain spatial derivatives of the fields
$C_i$, e.g. a Laplacian term representing diffusion.

Let us assume that the
advective transport dominates and diffusion is weak.
If we neglect non-local processes like  
diffusion 
on the right hand side of (\ref{general}) 
$F_i$ 
becomes a simple function of the local concentrations and coordinates.
In this case the Lagrangian
representation leads to a considerable simplification of the
equations, leading to a
low-dimensional dynamical system
\begin{equation}
\frac{d{\bf r}}{dt}={\bf v}({\bf r},t),\;\; 
\frac{d{C_i}}{dt}={F_i}(C_1,C_2,..,C_N,{\bf r}),\;
i=1,.,N,
\label{dynamics}
\end{equation}
where the second set of equations describes 
the chemical dynamics inside a fluid parcel
that is advected by the flow according to the first equation.
The coupling between the flow and chemical evolution is non-trivial if 
some of the functions ${F_i}$ 
depend explicitly 
on the coordinate ${\bf r}$. 
In applications this dependence of $F_i$ can appear as a consequence of
spatially varying sources, spatially varying (e.g. temperature dependent)
reaction rates or reproduction rates of biological species. It is
therefore natural to include this dependence in the model to be considered.

Although we use a Lagrangian representation, our aim is to follow not
the chemical evolution along individual trajectories, but the
spatiotemporal
dynamics of the chemical fields $C_i({\bf r},t)$ that is equivalent to the
evolution of the ensemble of fluid parcels under the dynamics (\ref{dynamics}).
Thus the original problem defined in an infinite dimensional phase
space is reduced to an ensemble problem in a low-dimensional dynamical system.

By neglecting diffusion, we may miss certain classes of behaviour such as
propagating waves, typical for reaction-diffusion systems \cite{Kuramoto}.
However, as we shall see later, we capture a
non-trivial behaviour that we believe to be characteristic of the
full advection-reaction-diffusion problem when diffusion is weak.

We assume that the flow ${\bf v}({\bf r},t)$ is two-dimensional incompressible
and periodic in time with period $T$.
These conditions in general
lead to chaotic advection even in case of simple spatially smooth
(non-turbulent) velocity fields. 
Since the advection is independent of the chemical dynamics it can be
characterised by its own Lyapunov exponents $\lambda^F_1>\lambda^F_2$. 
Incompressibility
implies 
that the advection is described by a conservative dynamical system
and thus
$\lambda^F_1=-\lambda^F_2$.

For numerical investigations we used a simple time-periodic flow
that consists of the alternation of two steady sinusoidal
shear flows in the $x$ and $y$ direction for the first and the
second half of the period, respectively. The flow is defined on the
unit square with periodic boundary conditions by the velocity field
\begin{eqnarray}
v_x(x,y,t)& =& -\frac{U}{T} \Theta \Bigl (\frac{T}{2}-t \bmod T \Bigr )
 \cos({2\pi y}) \nonumber \\
v_y(x,y,t)&=& -\frac{U}{T} \Theta \Bigl (t \bmod T-\frac{T}{2} \Bigr )
\cos({2\pi x})
\label{flow}
\end{eqnarray}
where $\Theta(x)$ is the Heavyside step function.
$U \to 0$ corresponds to an integrable limit of the advection problem.
We will consider the case $U=1.0$ producing a flow that consists
of one connected chaotic region.
The parameter $T$ controls the relative time-scale of the flow and reactions.
By changing $T$ we can vary
the Lyapunov exponents of the flow without altering the 
shape of the trajectories and the
spatial structure of the flow.

Since the trajectories ${\bf r}(t)$ are chaotic
the chemical dynamics (\ref{dynamics})
corresponds to a chaotically driven dynamical system.
This subsystem can be characterised by the set of chemical Lyapunov exponents
$\lambda^C_1>\lambda^C_2>...>\lambda^C_N$, 
that depend on the driving by the chaotic advection.
Here we will only consider the simplest situation, 
when the largest chemical
Lyapunov exponent is negative and, for a fixed trajectory ${\bf r}(t)$, 
the chemical evolutions 
converge to the same globally attracting chaotic orbit for any initial condition
in the chemical subspace.
In the case of no explicit space dependence in the functions 
$F_i$ 
(i.e. no driving) this would correspond 
to a globally attracting fixed point of the
chemical subsystem.

A simple example of such a system (relevant for atmospheric 
photochemistry) is the decay of a
chemical species ($N=1$) produced by a non-homogeneous source, 
with chemical dynamics 
\begin{equation}
{\dot C} = a({\bf r})-b C,
\label{decay}
\end{equation}
for which $\lambda^C= -b$. 

First we investigate the temporal evolution of the chemical fields.
Although the Lagrangian variables $C_i(t)$ have a chaotic time-dependence
according to the positivity of the largest Lyapunov exponent for
the full dynamical system (\ref{dynamics}), 
it can be shown that the Eulerian chemical field $C_i({\bf r},t)$
is asymptotically periodic in time with the period of the flow.
In order to obtain the values of the fields at point ${\bf r}$ at time $t$ one
can $i)$ integrate the advection 
backward in time and use the obtained trajectory ${\bf r}(t')$ ($0<t'<t$) 
and the
initial values of the chemical fields at the point ${\bf r}(0)$
to $ii)$ integrate the chemical dynamics forward in time from 0 to $t$.
The value of the field at the same point at time $t+nT$ can be obtained
similarly. The resulting backward trajectory will be the same (due to the
periodicity of the flow) but longer. By integrating the chemical
evolution forward in time for $nT$ we obtain a problem equivalent to the
previous one with a
different set of 
initial concentrations, that according to the assumptions made above
($\lambda^C_1<0$)
converge to the same orbit.
Consequently the chemical fields
are asymptotically periodic in time 
\begin{equation}
\lim_{m \to \infty} C_i({\bf r},(m+n)T+\tau)=\lim_{m \to \infty} C_i({\bf r},
mT+\tau),
\end{equation}
where $\tau = t \bmod T$, defining an asymptotic chemical field 
$C^{\infty}({\bf r},\tau)$.

In the following we investigate the spatial structure
of the chemical fields.
For this we calculate the difference 
\begin{equation}
\delta C_i= C_i({\bf r}+\delta {\bf r},t)-C_i({\bf r},t),\;\;\;\;
 \;\;\;|\delta {\bf r}| \ll 1, 
\label{difference1}
\end{equation}
that can be obtained by integrating (\ref{dynamics}) along two trajectories 
ending at
the preselected points ${\bf r}$ and ${\bf r}+\delta {\bf r}$ at 
time $t$.  
The time evolution of the distance  $|\delta {\bf r}(t')|$
(for $t-t' \gg 1$) can be estimated from the
time reversed advection dynamics as
\begin{equation}
|{\bf \delta r}(t')|\sim|{\bf \delta r}(t)| e^{\lambda^F(t-t')}
\label{dispersion}
\end{equation}
where $\lambda^F=\lambda^F_1>0$ for almost all 
final orientations 
${\bf n}(t) \equiv \delta {\bf r}(t)/|\delta {\bf r}(t)|$.
The only exception is the unstable contracting direction of the
time reversed flow corresponding to $\lambda^F=\lambda^F_2<0$.

By expanding (\ref{dynamics}) around the chaotic orbit 
$C_i({\bf r}(t))$ we obtain
a set of linear equations 
\begin{equation}
\dot {\delta {C_i}}= \sum_{j=1}^N {\partial F_i\over \partial {C_j}}
\delta {C_j}+
{\nabla_r {F_i} }\cdot {\delta \bf r}
\label{linear}
\end{equation}
with initial condition
\begin{equation}
\delta {C_i}(0)= C_i^0({\bf r}(0)+{\bf \delta r}(0))-C_i^0({\bf r}(0)),
\end{equation}
where $C_i^0({\bf r})$ are the initial chemical fields.

In the simplest case $N=1$ the solution of (\ref{linear}) can be 
written explicitly as
\begin{equation}
\delta {C}(t)=\nabla_r {C^0} \cdot \delta {\bf r}(0)
e^{\lambda t}+\int_0^t \nabla_r {F} \cdot \delta {\bf r}(t')
e^{\lambda (t-t')} dt'
\label{difference}
\end{equation}
where 
\begin{equation}
\lambda={1 \over t} \int_0^t 
{\partial F \over \partial C}(C({\bf r}(t')),{\bf r}(t'))dt'
\end{equation}
that becomes equal to the chemical Lyapunov exponent $\lambda^C$ 
in the $t \to \infty$ limit.
The first term in (\ref{difference}) represents a deviation due to 
non-homogeneous initial conditions and the second one describes the
effect of different histories of the two trajectories appearing via the
space dependence of $F$.

Using (\ref{dispersion}) we obtain from (\ref{difference}) for 
the projection of the gradient of $C$ to a direction ${\bf n}$ 
\begin{eqnarray}
{\delta C(t) \over |\delta {\bf r}(t)|}&=&
\nabla C^0 {\bf n}(0)
e^{(\lambda^F+\lambda^C)t}+ \nonumber \\ 
&+&\int_0^t \nabla F({\bf r}(t')) {\bf n}(t')
e^{(\lambda^F+\lambda^C)(t-t')} dt'.
\end{eqnarray}
The convergence of the right hand side 
of the above equation for $t \to \infty$ depends on
the sign of the exponent $\lambda^F + \lambda^C$.
If $\lambda^F < |\lambda^C|$ the convergence of the chemical dynamics is
stronger than the dispersion of the trajectories 
in the physical space 
and results in a smooth field $C^{\infty}({\bf r},\tau)$. On the contrary,
if $\lambda^F > |\lambda^C|$ the memory of the chemical dynamics 
decays too slowly to forget the different spatial histories (or initial
conditions). In this case the limit does not exist and
the field $C^{\infty}({\bf r},\tau)$ has an irregular
structure that is almost nowhere differentiable. 
There exists, however, at each point one special direction in which 
the derivative is finite. This direction is the contracting
direction in
the time-reversed advection dynamics 
corresponding to the negative Lyapunov exponent of the
flow $\lambda^F_2$.
Thus we suggest a precise definition of a {\em filamental} structure, being a
non-differentiable field, that is
still smooth in one direction at each point (with that direction
itself varying smoothly).  

The irregular chemical field can be characterised by its
H\"older exponent $\alpha$ defined as
$\delta C(\delta {\bf r}) \sim |\delta {\bf r}|^{\alpha}$,
where $0<\alpha<1$ and $\alpha=1$ for a differentiable function.
In our case if $\lambda^F+\lambda^C>0$ 
\begin{equation}
\delta C(t) \sim |\delta {\bf r}(t)| e^{(\lambda^F+\lambda^C)t}.
\label{divergence}
\end{equation}
Expressing time from (\ref{dispersion}) as
\begin{equation}
t={1 \over \lambda^F}\ln{ |\delta {\bf r}(0)| \over |\delta {\bf r}(t)|}
\end{equation}
and inserting it in (\ref{divergence})
we obtain
\begin{equation}
\delta C(t) \sim |\delta {\bf r}(0)|^{(1+\lambda^C/\lambda^F)} |\delta {\bf r}(t)|^
{-\lambda^C/\lambda^F}. 
\label{holder}
\end{equation}
For very long times and in a closed flow
$|\delta {\bf r}(0)|$ will saturate at a finite 
value in the backward advection dynamics.
Thus the H\"older exponent is $\alpha=|\lambda^C|/\lambda^F$.
Certainly, diffusion would smooth out the small
scale filamental structures below a certain diffusive
scale (approaching zero for smaller and smaller diffusivities),
setting a cut-off for the scaling relation (\ref{holder}).
Nevertheless, above the diffusive scale filamental structures will
persist for arbitrarily long time since,
in the presence of chemical sources, the effect of diffusion is
balanced by the continuous generation of small scale
structures by the chaotic advection. 

As an example we consider the system (\ref{decay}),
which is the simplest possible system that exhibits 
the smooth-filamental transition, using a source 
term of the form $a(x,y)=1+a_1 \sin{(2 \pi x)}\sin{(2 \pi y)}$.
Numerically calculated chemical fields (obtained by a backward integration of
the advection problem 
and forward integration of the chemical 
dynamics 
\cite{simulation}) 
of both types are shown in Fig.1a and b. Figure 2 shows
a section of the smooth and filamental fields of Fig.1 along the line 
$y=0.25$. 
We also measured the box counting fractal dimension of the function $C({\bf r})$
along a cut in the $x$ direction, which is related to the H\"older exponent by 
$D=2-\alpha$ \cite{Turcotte}. Numerically computed values agree well with 
the theoretical prediction as shown in Fig. 3.
If the flow does not consist of only one connected chaotic region, 
but is composed by 
chaotic regions separated by KAM tori, 
the 
Lyapunov exponents of the flow are different in each chaotic region
and the structure of the chemical field can be
smooth in certain regions but filamentary in others (Fig 1c.).

In the more general case $N>1$ the same smooth-filamental
transition can be obtained. 
Let us once again consider at a particular time $t$ the difference
$\delta C_i(t)$ in the chemical fields over a small preselected 
displacement $\delta {\bf r}$.
The time dependence of a set of initial deviations
$\delta C_i(0)$ and $\delta {\bf r}(0)$,
for the dynamical system (\ref{dynamics}) has the asymptotic form
\begin{equation}
\Bigl (\sum_{i=1}^N \delta C_i^2(t)+\delta {\bf r}^2(t)\Bigr )^{1/2} \sim 
\Bigl (\sum_{i=1}^N \delta C_i^2(0)+\delta {\bf r}^2(0) \Bigr )^{1/2} e^{\lambda t}
\label{Lyapunov}
\end{equation}
where $\lambda$ is one of the Lyapunov exponents of the system.
For a typical choice of the final deviation $\delta {\bf r}(t)$, 
$\delta {\bf r}$
will be divergent in the time reversed dynamics, and consequently 
contracting in the forward direction.
Thus, the contribution of $\delta {\bf r}$ decays in the left hand side of 
(\ref{Lyapunov})
and $\lambda$ cannot be the positive Lyapunov exponent $\lambda^F_1$
so the typical value of $\lambda$ will be the second largest 
Lyapunov exponent. 
There are thus two possibilities $\lambda=\lambda^F_2=
-\lambda^F_1$ (if $\lambda^F_2>\lambda^C_1$) or $\lambda=\lambda^C_1$
(otherwise).
We can now divide both sides of (\ref{Lyapunov}) by $\delta {\bf r}(t)$ 
using (\ref{dispersion})
\begin{equation}
\Bigl (\sum_{i=1}^N {\delta C_i^2(t) \over \delta {\bf r}^2(t)}+1\Bigr )^{1/2} \sim
\Bigl (\sum_{i=1}^N {\delta C_i^2(0) \over \delta {\bf r}^2(0)}+1 \Bigr )^{1/2}
e^{(\lambda^F_1+\lambda)t}.
\end{equation}
Thus, the 
chemical fields $C_i({\bf r})$ become non-differentiable in the $t \to \infty$
limit if $\lambda^F_1>|\lambda^C_1|$.

Numerically we have also observed the smooth-filamental transition 
in some two-component systems (Brusselator autocatalytic reaction model 
\cite{Brussel},
phytoplankton-zooplankton population dynamics model \cite{plankton}) for
parameters that satisfies the condition $\lambda^C_1<0$, and obtained
distributions very similar to that shown in Fig.1.

Useful discussions with B. Fernandez,
E. Hern\'andez-Garc\'\i a and J. Vanneste are acknowledged.
Z.N. was funded by a Royal Society-NATO Postdoctoral Fellowship and
C.L. was funded by CICYT projects (MAR95-1861) and (MAR98-0840).

\begin{figure}
\caption{
Smooth $(a)$ and filamental $(b)$ distributions
of the decaying substance (eq.\ref{decay}) mixed by the
flow (eq. \ref{flow}) after 20
periods 
using $a_1=0.1$ ,$b=1.0$ ,$U=1.0$ 
and initial condition $C({\bf r},t=0)=0.0$.
The integration has been done for 
$200 \times 200$ points with final positions on a rectangular grid.
For $U=1.0$ the 
numerically calculated Lyapunov exponent of
the flow is $\lambda^F \approx 2.35/T$. The period of
the flow is $T=5.0$ ($\lambda^F<b$) in $(a)$ 
and $T=1.0$ ($\lambda^F>b$) in $(b)$, respectively.
$(c)$ coexistence of smooth and filamental structures
for $U=0.5$ and $T=1.0$.
}
\end{figure}
\begin{figure}
\caption{
Sections of the smooth $(a)$ and filamental $(b)$ fields
shown in Fig.1 along the line $y=0.25$.
}
\end{figure}
\begin{figure}
\caption{
Box counting fractal dimensions of the functions $C(x)$
shown in Fig.2.
Number of boxes needed to cover the graph of the
function $C(x)$ vs. the box size $l$
(squares) and slopes corresponding to
the relation $D=2-b/\lambda^F$. 
}
\end{figure}
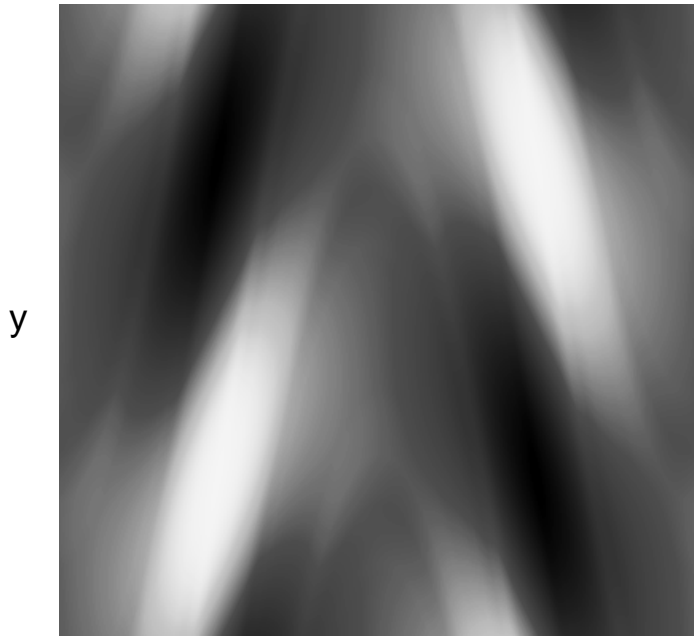
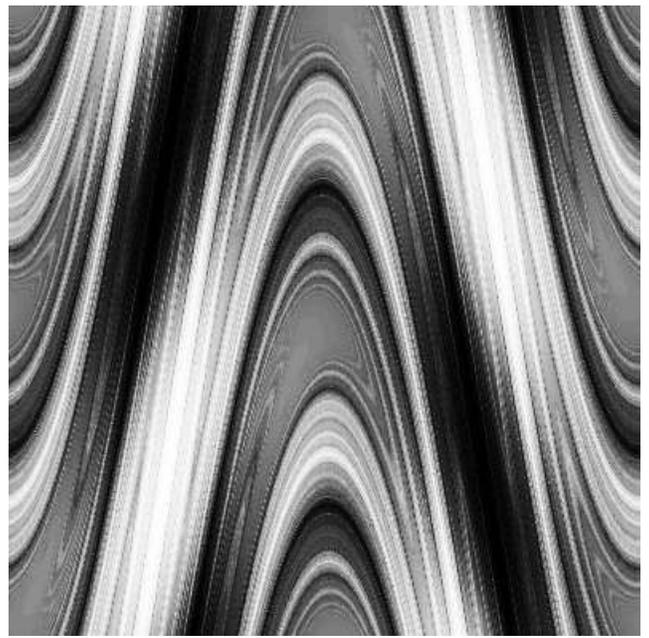
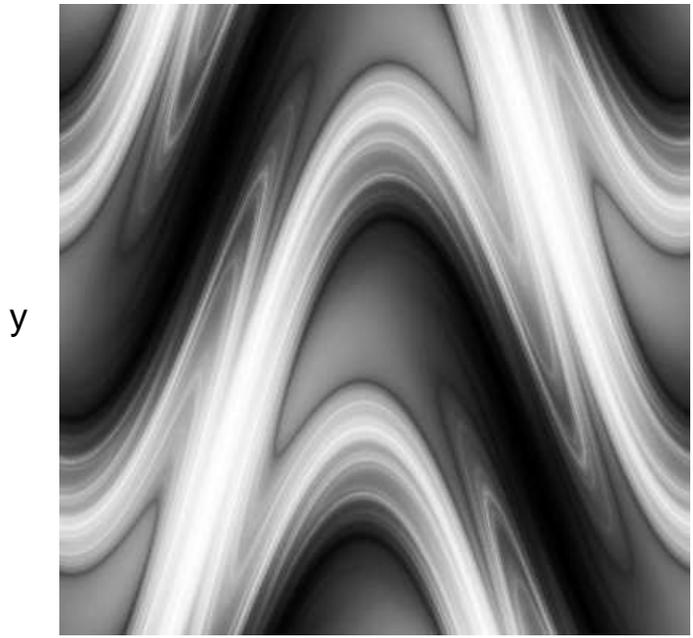
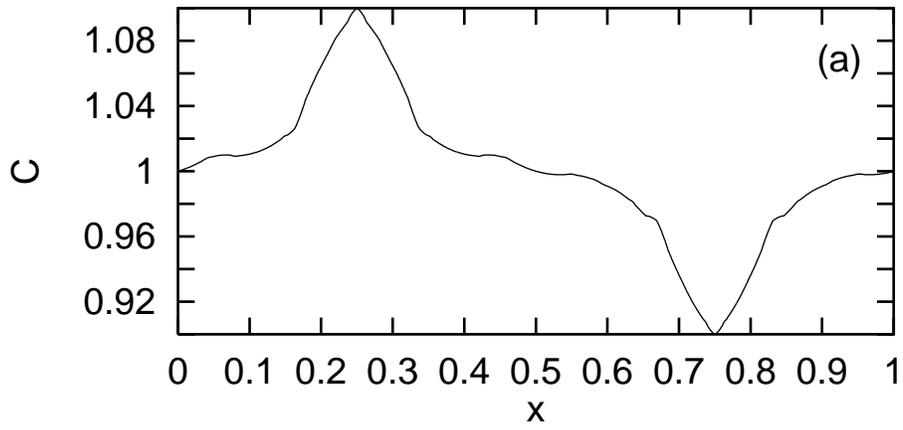
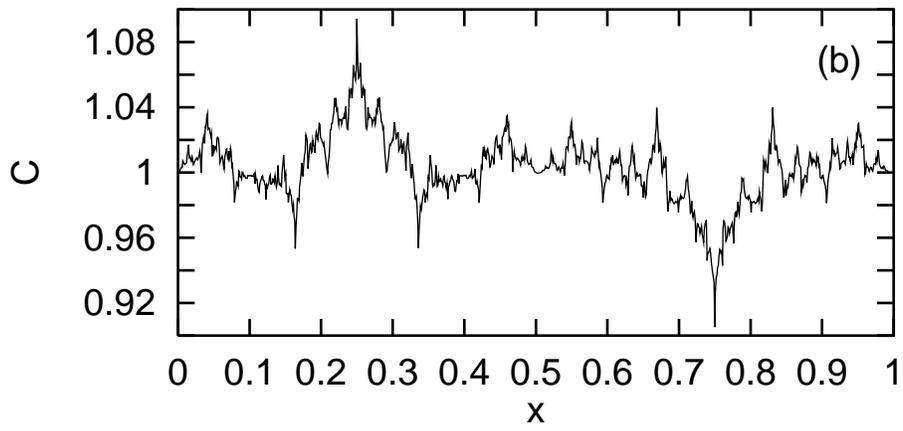
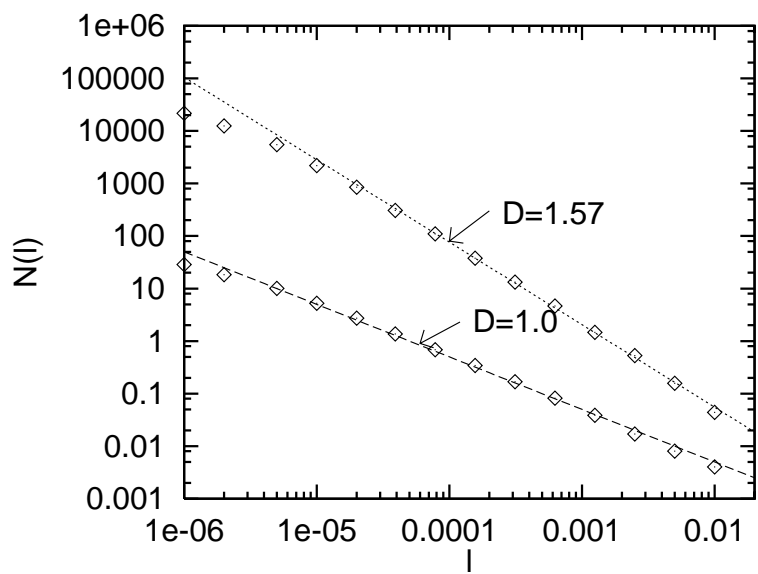
\end{document}